\documentclass[12pt,twoside]{article}
\usepackage{fleqn,espcrc1}

\usepackage{epsfig}
\usepackage{rotating}

\newcommand{\AmS}{{\protect\the\textfont2
  A\kern-.1667em\lower.5ex\hbox{M}\kern-.125emS}}

\title
{The  NLO DGLAP extraction of $\alpha_s$ and higher twist 
terms from CCFR $xF_3$ and $F_2$ structure functions: results and scale 
dependence}

\author{        S.I. Alekhin\address{Institute for High Energy Physics,
        142284 Protvino, Russia}
        and
        A.L. Kataev\address{Institute for Nuclear Research of the
        Academy of Sciences of Russia,\\
        117312 Moscow, Russia}
        \thanks{Supported in part by the Russian Foundation
        of Basic Research, Grant N 99-01-00091}}
\begin{document}

\maketitle

\begin{abstract}
We perform a detailed NLO analysis of the combined CCFR $xF_3$ 
and $F_2$ structure functions data and extract the value of 
$\alpha_s$, parameters of distributions and higher-twist (HT) terms 
using a direct solution of the DGLAP equation. The value 
of $\alpha_s(M_Z)=0.1222\pm 0.0048 $(exp)$\pm 0.0026$(theor) is obtained.
Our result has a larger central value and errors than the original 
one of the CCFR collaboration due to model independent parametrization of
the HT contributions. The dependence of HT contributions on the 
QCD renormalization scale is studied.
\end{abstract}

\vskip 1cm
In the recent years interest to the 
problem of the extraction of the high-twist-terms from the 
analysis of different deep-inelastic scattering (DIS) data was renewed, mainly due to the 
possibility to model these terms in different processes using the
infrared-renormalon (IRR) technique   
(see e.g. Refs. \cite{BB}-\cite{AZ} 
and, especially, Ref. \cite{Beneke} for the review).

On the other hand, the experimentalists improve 
their data precision and achieve, sometimes, a percent level 
of accuracy. For example, very precise data on $xF_3$ and $F_2$  
from the $\nu N$ DIS experiment, 
performed at Tevatron by the CCFR collaboration, recently appeared
\cite{CCFR,SELIG}. The CCFR data on $xF_3$ were 
analyzed in Ref. \cite{KKPS} in  the next-to-leading-order (NLO), 
and with an approximate next-to-next-to-leading order (NNLO) corrections. 
For the latter the  NNLO QCD 
corrections  to the coefficient function \cite{VZ} were taken into account. 
The  NNLO corrections to the  
anomalous dimensions of a limited set of even non-singlet moments 
\cite{LRV} were also taken into account. 
The NNLO corrections to the anomalous dimensions of odd moments, 
which are not still explicitly calculated, were obtained using
smooth interpolation procedure
proposed in Ref.  \cite{PKK} and improved in Ref. \cite{KKPS}.
The aim of Ref.  \cite{KKPS} was to attempt the first NNLO  determination of  
$\alpha_s(M_Z)$ from  DIS and to extract the HT terms from the data 
on $xF_3$ within the framework of the IRR-model \cite{DW}.
Alongside, the model-independent extraction of the HT terms was made, 
similarly to the analysis 
of the combined SLAC-BCDMS data \cite{VM}, 
which was performed in the NLO approximation.
Theoretical uncertainties of the analysis of Ref. \cite{KKPS} 
were further estimated 
in Refs. \cite{KPS} in the N$^3$LO approximation  
using the method of Pad\'e approximants. 
It has been found in Refs. \cite{KKPS,KPS} that the inclusion of
the NNLO corrections leads to the decrease
of the HT terms, so that at the  NNLO its $x$-shape variation 
is closer to zero.

In this work we completed the fits of  Ref.\cite{AK}, performed the 
NLO analysis of the CCFR data on the structure functions $F_2$ and $F_3$ 
with the help of a QCD DGLAP evolution code, developed in Ref. \cite{ALEK}. 
It should be stressed that  
the code \cite{ALEK} was tested using the procedure 
proposed in Ref. \cite{BENCH} and demonstrated the accuracy at the level of 
$O(0.1\%)$ in the kinematic region covered by the analyzed data.
Our fits were made in the NLO approximation within the 
modified-minimal-subtraction
($\overline{MS}$) factorization and renormalization schemes.
The effective number of flavours $n_f$
was chosen to be $n_f=4$ for $Q^2$ less than the definite scale $M_5^2$ 
and increased to $n_f=5$ at larger values of $Q^2$
keeping the continuity of $\alpha_s$ \cite{BW}.
The value of the effective matching scale $M_5$ was varied from 
$M_5=m_b$ to $M_5=6.5m_b$. The latter  choice
was advocated in Ref. \cite{BN} on the basis of the DIS sum rules consideration. 
The dependence of the fit results on the choice of the matching point 
gives one of the sources of theoretical uncertainties inherent 
to our analysis. 

The leading twist terms $xF_3^{LT}(x,Q)$ and $F_2^{LT}(x,Q)$ were obtained 
by direct integration of the DGLAP equations for non-singlet, pure-singlet, and 
gluon distributions that were 
subsequently convoluted with the coefficient functions.
In order to provide the straightforward way for comparison of our
results with Ref. \cite{ALEK}, the
initial reference scale for pQCD evolution $Q_0^2=9~$GeV$^2$ was taken. 
The boundary conditions at this reference scale 
were chosen in the form 
analogous to the ones, used in Refs. \cite{SELIG,KKPS}:
\begin{displaymath}  
xq^{NS}(x,Q_0)=\eta_{NS}x^{b_{NS}}(1-x)^{c_{NS}}(1+\gamma x)\frac{3}{A_{NS}}
\end{displaymath}
for non-singlet distribution,
\begin{displaymath}  
xq^{PS}(x,Q_0)=\eta_{S}x^{b_S}(1-x)^{c_S}/A_{S}
\end{displaymath}  
for pure-singlet distribution, and
\begin{displaymath}  
xG(x,Q_0)=\eta_{G}x^{b_G}(1-x)^{c_G}/A_{G}
\end{displaymath}  
for gluon distribution,
where $A_{NS}$, $A_{S}$, and $A_{G}$ were defined from 
the partons' number/momentum conservation and other parameters were
fitted.

The expression for the $xF_3$ and $F_2$ that includes the 
HT contribution looks as follows:
\begin{displaymath}  
xF_3^{HT}(x,Q)=xF_3^{LT,TMC}(x,Q)+\frac{H_3(x)}{Q^2},~~~ 
F_2^{HT}(x,Q)=F_2^{LT,TMC}(x,Q)+\frac{H_2(x)}{Q^2},
\end{displaymath}  
where $F_{2,3}^{LT,TMC}(x,Q)$ are $F_{2,3}^{LT}(x,Q)$
with the target mass correction  applied. 
We used  the model independent HT-expression, i.e.
$H_{2,3}(x)$ were parametrized
at $x=0.,0.2,$ $0.4,0.6,0.8$ with linear interpolation 
between these points. 

 The account of the point-to-point correlations of the data due to 
systematic errors can be crucial for the estimation of total 
experimental errors of the 
parameters (see, in particular, Ref. \cite{Aln}, where the value
$\alpha_s(M_Z)=0.1180\pm 0.0017~$(stat+syst) was obtained as a 
result of the combined fit to the SLAC-BCDMS data with HT included). 
The systematic errors were taken into account
analogously to the  works of Refs.\cite{ALEK,Aln}.
The total number of the independent 
systematic errors sources for the analyzed data is 18 and all of them  
were convoluted
into a general correlation matrix, which was used for the construction of
the minimized $\chi^2$. 
In addition to the point-to-point correlation of the data due to
systematic errors, the statistical correlations between $F_2$ and $xF_3$
were also taken into account. The  account of 
systematic errors leads to a significant increase of the 
experimental uncertainties  of the HT parameters and  
the shift of their central values.
However, even in this case, there is a definite  
agreement with the results on HT-behaviour of Ref. \cite{KKPS},
obtained  in NLO. Moreover, these results do not contradict the 
IRR-model prediction of Ref. \cite{DW}, since fitting IRR model 
parameter $A_2^{'}$ to the data, we obtained $A_2^{'}=-0.10 \pm 0.09$~GeV$^2$, 
compatible with $A_2^{'}\approx-0.20$~GeV$^2$ and 
$A_2^{'}\approx -0.1~GeV^2$, given in Ref. \cite{DW} and Ref.\cite{Stein}.

\begin{figure}[t]
\centerline{\epsfig{file=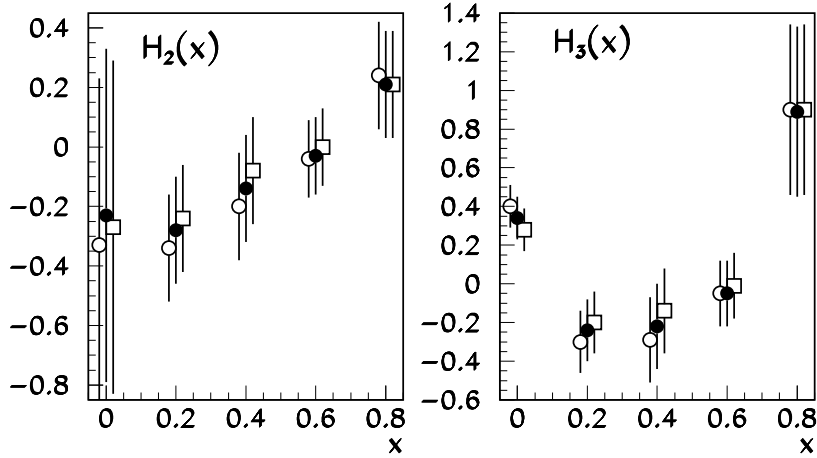,height=6cm}}
\caption{The high-twist contribution to the structure functions $F_2$ $F_3$.
Full circles correspond to the fit with renormalization scale parameter
$k_R=1$, empty circles -- to the fit with $k_R=1/4$,
squares -- to the fit with $k_R=4$.}
\end{figure}

Performing the trial fits we got convinced that 
the introduction of the factor $(1+\gamma x)$ 
into the reference expressions for the
the gluon and singlet distributions does not improve the quality of the fit
and does not change the value of $\alpha_s$.
Also, we fixed parameters $\gamma_{NS}, b_S$ and $b_G$ 
at zero because this increase the value of $\chi^2$ by few units only 
while $\chi^2/$NDP remained less than unity. 

The results of the 
fit on $H_2(x)$ and $H_3(x)$ parameters are given in Fig. 1. 
One can notice that, comparing with the fit to $xF_3$ data only
from Ref. \cite{KKPS}, the HT parameters errors are decreasing.
Within the errors, the parameters that describe   
the boundary distributions are compatible with ones of Ref. \cite{SELIG}.
The $H_3(x)$ coefficients  are in
agreement with the NLO results of Ref. \cite{KKPS}
and the behaviour of $H_2(x)$ qualitatively reproduce the 
HT contribution to $F_2$ that was obtained 
from the combined fits to the SLAC-BCDMS data on $F_2$ \cite{VM,Aln}.

When the matching scale $M_5$ was changed from $m_b$ to $6.5m_b$,
the value of $\alpha_s(M_Z)$ shifted down by 0.0052 and, hence,
the theoretical error in $\alpha_s(M_Z)$ due to uncertainty 
of b-quark threshold can be estimated as 0.0026. 
This uncertainty is in agreement with the results of the 
fits to the CCFR data obtained within the 
so-called spline $\overline{MS}$ prescription \cite{ShSM}
with the help of the Jacobi polynomial metod \cite{Jacobi}.
One more source of the theoretical uncertainty
due to the truncation of higher QCD orders was evaluated
following the way, which was proposed in Ref. \cite{VM}. 
In accordance with their procedure,
one can  introduce renormalization scale $k_R$
into QCD evolution equations in the way, given below 
for non-singlet evolution:
\begin{displaymath}
\frac{dxq^{NS}}{d\ln Q}=\frac{\alpha_s(k_RQ)}{\pi}\int^1_x
dz\biggl\{P^{NS,(0)}_{qq}(z)+
\end{displaymath}
\begin{displaymath}
+\frac{\alpha_s(k_RQ)}{2\pi}
\Bigl[P^{NS,(1)}_{qq}(z)+\beta_0P^{NS,(0)}_{qq}(z)\ln(k_R)\Bigr]
\biggr\}\frac{x}{z}q^{NS}(x/z,Q),
\end{displaymath}
where $P^{NS,(0)}$ and $P^{NS,(1)}$ denote the  LO and the NLO parts of 
the non-singlet splitting function.
The dependence of the results on $k_R$ would signal an incomplete 
account of the perturbation series.
The shift of $\alpha_s(M_Z)$ 
resulting from the variation of $k_R$ from 1/4 to 4 
turned out to be only 0.0007. At the same time one can observe
a simultaneous variation of $H_{2,3}(x)$ (see Fig. 1). This effect can 
denote,
that, in fact, the fitted values of $H_{2,3}(x)$ incorporate 
the higher order QCD contributions (NNLO and beyond, c.f. 
Ref. \cite{KKPS,KPS}). 
It is interesting, that the same effect was also observed in the analysis
of charged leptons data \cite{AlekhinM}, 
i.e. it cannot be attributed to a specific feature of the CCFR data. 
The interplay between these contributions and genuine power corrections does
not allow for their unambiguous separation. The 
value of $\alpha_s$ is strongly correlated with the fitted high
twist, that leads to increase of the $\alpha_s$ error and,
consequently, in our fit with simultaneous determination 
of $\alpha_s$ and high twist contributions the 
theoretical error in $\alpha_s$
due to truncation of higher QCD orders is merged into 
the total experimental errors. 
Having taken $Q_0^2=20~$GeV$^2$ as an initial scale, we
checked that our NLO results  were quite stable to the variation 
of the initial scale. However, we do not know what will 
happen at the NNLO, where the stability to $Q_0^2$ was observed 
only starting from $Q_0^2\approx 20~GeV^2$ \cite{KPS}.   

The final value of $\alpha_s$ in NLO with the account of 
theoretical uncertainties is given as 
$\alpha_s(M_Z)=0.1222\pm 0.0048~{\rm (stat+syst)} 
\pm 0.0026~{\rm (theor.)}$ 
It  differs a bit from the NLO  value 
$\alpha_s(M_Z)=0.119 \pm 0.002~$(stat+syst)$\pm0.004$~(theory)
obtained in the CCFR analysis \cite{CCFR}. The increase of the
experimental error is due to that CCFR group
used model-dependent form of the HT contributions,
while we considered them as
the additional free parameters and extracted them from the fit.  

It should be stressed, that the scale-dependence uncertainty of the NLO results 
drastically minimized to the value of 0.0007
after taking into account HT corrections. The decrease 
of this uncertainty was also recently found 
in the analysis without HT terms 
after taking into account 3-loop splitting function \cite{NV}. 
Thus we think that more careful NNLO DGLAP analysis of the DIS 
data with HT effects included is now on the agenda.

{\bf Acknowledgements} One of us (ALK) would like to thank 
the Organizers of Nucleon99 Workshop for hospitality in Frascati 
and for the financial support.

\end{document}